\documentclass[secnumarabic,amssymb,nobibnotes,nofootinbib,aps,prc]{revtex4}
\usepackage{psfig}
\usepackage{graphicx}
\usepackage{bm}
\newcommand{\bea}{\begin{eqnarray}}
\newcommand{\eea}{\end{eqnarray}}
\newcommand{\be}{\begin{equation}}
\newcommand{\ee}{\end{equation}}
\newcommand{\bt}{\begin{tabular}}
\newcommand{\et}{\end{tabular}}
\newcommand{\Tr}{{\rm Tr}}
\newcommand{\no}{\nonumber}
\newcommand{\ovl}{\overline}

\newcommand{\Si}{ \mbox{\boldmath $\Sigma$}  }

\newcommand{\pa}{\partial}
\newcommand{\beas}{\begin{eqnarray*}}
\newcommand{\eeas}{\end{eqnarray*}}

\newcommand{\fr}{\frac}

\newcommand{\dg}{\dagger}
\newcommand{\La}{\Lambda}
\newcommand{\pam}{\partial_\mu}
\begin{document}
\baselineskip 6.5mm
\date{\today}
\title{Mass modification of $D$-meson in hot hadronic matter}
\author{A.Mishra}
\email{mishra@th.physik.uni-frankfurt.de}
\affiliation{Institut f\"ur Theoretische Physik,
        Robert Mayer Str. 8-10, D-60054 Frankfurt am Main, Germany}

\author{E.L. Bratkovskaya}
\affiliation{Institut f\"ur Theoretische Physik,
        Robert Mayer Str. 8-10, D-60054 Frankfurt am Main, Germany}

\author{J. Schaffner-Bielich}
\affiliation{Institut f\"ur Theoretische Physik,
        Robert Mayer Str. 8-10, D-60054 Frankfurt am Main, Germany}

\author{S. Schramm}
\affiliation{Institut f\"ur Theoretische Physik,
        Robert Mayer Str. 8-10, D-60054 Frankfurt am Main, Germany}

\author{H.~St\"ocker}
\affiliation{Institut f\"ur Theoretische Physik,
        Robert Mayer Str. 8-10, D-60054 Frankfurt am Main, Germany}

\begin{abstract}
We evaluate the in-medium $D$ and $\bar D$-meson masses in hot hadronic
matter induced by interactions with the light hadron sector described
in a chiral SU(3) model. The effective Lagrangian approach is generalized 
to SU(4) to include charmed mesons. We find that the D-mass drops 
substantially at finite temperatures and densities, which open the
channels of the decay of the charmonium states
($\Psi^\prime$, $\chi_c$, $J/\Psi$) 
to $D \bar D$ pairs in the thermal medium. 
The effects of vacuum polarisations from the baryon sector on the 
medium modification of the $D$-meson mass relative to those obtained 
in the mean field approximation are investigated.
The results of the present work are compared to
calculations based on the QCD sum-rule approach, 
the quark-meson coupling model, chiral perturbation theory,
as well as to studies of quarkonium dissociation
using heavy quark potential from lattice QCD.

\end{abstract}
\maketitle

\def\bfm#1{\mbox{\boldmath $#1$}}

\section{Introduction}
The in-medium properties of hadrons have been a field of active
theoretical  research in the last decade. From the experimental side
in-medium effects have also been probed in relativistic heavy-ion
collisions from SIS (1-2 A$\cdot$GeV) to SPS energies (30-160
A$\cdot$GeV).  At the SPS the experimentally observed dilepton spectra
\cite{ceres,helios} -- that offer a direct glance at the in medium
spectral properties of the vector mesons -- have been attributed to
medium modifications of the spectral function especially of the
$\rho$-meson \cite{Brat1,CB99,vecmass,dilepton,liko} and can not be
explained by vacuum hadronic properties.

Furthermore, experiments on $K^\pm$ production from nucleus-nucleus
collisions at SIS energies (1-- 2 A$\cdot$ GeV) have shown that
in-medium properties of kaons are seen in the collective flow pattern
of $K^+$ mesons both, in-plane and out-of-plane, as well as in the
abundancy and spectra of antikaons
\cite{CB99,cmko,lix,Li2001,K5,K6,K4,kaosnew}.  Since the strangeness
sector ($K^{\pm}, \Lambda_s, \Sigma_s$) shows some analogy to the open
charm sector ($D^{\pm}, \Lambda_c, \Sigma_c$), which results from
exchanging strange/antistrange quarks by charm/anticharm quarks, the
medium modifications for the $D$-mesons have become a subject of recent
interest, too \cite{arata,liuko,friman,weise,digal,qmc}.  It is
expected that one might find open charm enhancement in nucleus-nucleus
collisions \cite{cassing} as well as J/$\Psi$ suppression
as experimentally observed at the SPS \cite{NA501}. Indeed, the NA50
Collaboration has claimed to see an open charm enhancement by up to a
factor of three in central $Pb+Pb$ collisions at 158 A$\cdot$GeV
\cite{NA50e}.

Open charm mesons ($D, D^*, \bar D, \bar{D^*}$) can be produced
abundantly in high energy heavy-ion collisions and might even dominate
the high mass (M $>$ 2 GeV) dilepton spectra \cite{lin}. The medium
modification of $D$-mesons is worth investigating since they should
modify the J/$\Psi$ absorption cross section in the medium with baryons
and mesons and could also provide a possible explanation for the
observed J/$\Psi$ suppression \cite{NA501}.  On the other hand, in high
energy heavy-ion collisions at RHIC ($\sqrt s \sim$ 200 GeV), an appreciable
contribution of J/$\Psi$ suppression is expected to be due to the
formation of a quark-gluon Plasma (QGP) \cite{blaiz}.  However, the
effect of the hadron absorption of J/$\Psi$'s is still not negligible
\cite{zhang,brat5,elena}. It is thus important to understand the
charmed meson interactions in the hadronic phase.

The $D$-meson mass modifications have been studied using the QCD
sum rule approach (QSR) \cite{arata}. It was found that due to the
presence of a light quark in the $D$-meson the mass modification of
$D$-mesons has a large contribution from the light quark condensates
\cite{weise}. The J/$\Psi$ - being a $c\bar{c}$ vector state - has a
dominant contribution from the gluon condensates. Accordingly, the
substantial medium modification of the light quark condensate -- as
compared to the gluon condensate -- is attributed to a larger drop of
the $D$-meson mass as compared to the J/$\Psi$ mass modification.
Alternatively, within a linear density approximation in the QSR, the
$DN$ interaction is seen to be more attractive than the J/$\Psi$-N
interaction \cite{arata}.  This leads to a larger decrease of the
$D$-meson mass than that of the charmonium (about 10 times larger at
nuclear matter density $\rho_0$). It is seen that the large mass shift
of the $D$-meson originates from the contribution of the $m_c \langle
\bar q q \rangle_N$ term in the operator product expansion. In the QMC
model, the contribution from the $m_c \langle \bar q q \rangle_N$ term
is represented by a quark-$\sigma$ meson coupling. The QMC model
predicts the mass shift of the $D$-meson to be of the order of 60 MeV at
nuclear matter density \cite{qmc}, which is very similar to the value
obtained in the QCD sum rule calculations of Ref. \cite{arata,weise}.
Furthermore, lattice calculations for heavy quark potentials at finite
temperature suggest a similar drop \cite{digal,lattice}.

A reduction of the $D$-mass has direct consequences for the production of
open charm \cite{cassing} as well as J/$\Psi$ suppression \cite{jpsi}.
Recently, the NA50 collaboration \cite{NA502} has reported a strong
(so called `anomalous') J/$\Psi$  suppression \cite{satz}
in Pb-Pb collision at 158 AGeV. A possible explanation
of the J/$\Psi$ suppression
\cite{arata} is the large difference in the mass shifts for D and
J/$\Psi$. As suggested by the p-A collision data \cite{pAdata},
a large part of the  observed J/$\Psi$ is produced from the excited states
$\chi _ c$ and $\Psi '$.  Thus an appreciable drop of the $D$-meson mass
could lead to the decay of these excited states ($\chi _ c$ or
$\Psi ^ \prime $) in the medium to $D\bar D$ final states
\cite{brat6} and might lead to a lower yield for J/$\Psi$ accordingly.
These effects
could be explored at the future accelerator facility at GSI \cite{gsi}.

In this work we study the medium modification of the masses of charmed
mesons (D$^\pm$) due to their interaction with the light hadron sector.
The medium modification of hadronic properties in hot and dense
hyperonic matter has been studied in a chiral $SU(3)$-flavor model in
Ref. \cite{kristof1}. We generalise the model to $SU(4)$-flavor to
include charm mesons. Thus knowing the interactions of the charmed
pesudoscalar mesons, we investigate their mass modification in the hot
and dense matter due to their interactions with light hadrons.  Within
the model, both D$^\pm$ experience a drop in the thermal medium due to
an attractive interaction via exchange of scalar mesons.  The D$^+$ has
a drop in the medium due to a vectorial (as well as vector $\omega$
exchange) term, whereas the $D^-$ has positive contributions from the
vector terms in the thermal medium. The effects on the $D$-meson mass 
arising from terms of the type $(\partial_\mu D^+)(\partial
^\mu D^-)$ are taken into consideration, too. We compare the results 
obtained in the model with those
obtained from the interactions arising from chiral perturbation
theory, where a drop of the $D^\pm$ mass is found due to an
attractive nucleon scalar interaction (the so called sigma term) while
the vectorial interaction is responsible for a drop (rise) for the
D$^+$ (D$^-$) mesons.  The repulsive scalar interaction ($\sim
(\partial_\mu D^+)(\partial^\mu D^-)$) which is of the same order as
the attractive scalar interaction in chiral model, is
also taken into consideration to study the in--medium $D$-meson mass.
In the present model, the effect is seen to be larger than that arising
from chiral perturbation theory. The effect of taking into account the
nucleon Dirac sea for the study of hadronic properties, and hence their
effect on $D$-mesons due to their interaction with the nucleons and
scalar and vector mesons, is studied additionally. It is seen to give
rise to higher values for the masses as compared to the mean field
approximation, since such vacuum polarizations effects decrease the
strength of the scalar and vector fields associated with a softening of
the equation of state.


We organize the paper as follows: We briefly recapitulate the
$SU(3)$-flavor chiral model adopted for the description of the hot and
dense hadronic matter in Section 2. The hadronic properties then are
studied within this approach. These give rise to medium modifications
for the $D$-masses through their interactions with the nucleons and scalar
and vector mesons as presented in Section 3.
Section 4 discusses the results of the present investigation, 
while we summarise our findings and discuss open questions in Section 5.

\section{ The hadronic chiral $SU(3) \times SU(3)$ model }
In this section the various terms of the effective Hadronic Lagrangian
\be
{\cal L} = {\cal L}_{kin} + \sum_{ W =X,Y,V,{\cal A},u }{\cal L}_{BW}
          + {\cal L}_{VP} + {\cal L}_{vec} + {\cal L}_0 + {\cal L}_{SB}
\label{genlag}
\ee
are discussed. Eq. (\ref{genlag}) corresponds to a relativistic quantum
field theoretical model of baryons and mesons built on chiral symmetry
and broken scale invariance \cite{paper3,hartree,kristof1} to describe
strongly interacting nuclear matter.  We adopt a nonlinear realization
of the chiral symmetry which allows for a simultaneous description of
hyperon potentials and properties of finite nuclei \cite{paper3}.  This
Lagrangian contains the baryon octet, the spin-0 and spin-1 meson
multiplets as the elementary degrees of freedom. In Eq. (\ref{genlag}),
$ {\cal L}_{kin} $ is the kinetic energy term, $  {\cal L}_{BW}  $
contains the baryon-meson interactions in which the baryon-spin-0 meson
interaction terms generate the baryon masses. $ {\cal L}_{VP} $
describes the interactions of vector mesons with the pseudoscalar
mesons (and with photons).  $ {\cal L}_{vec} $ describes the dynamical
mass generation of the vector mesons via couplings to the scalar
mesons and contains additionally quartic self-interactions of the
vector fields.  ${\cal L}_0 $ contains the meson-meson interaction terms
inducing the spontaneous breaking of chiral symmetry as well as
a scale invariance breaking logarithmic potential. $ {\cal L}_{SB} $
describes the explicit symmetry breaking of $ U(1)_A $, $ SU(3)_V $ and
the chiral symmetry.

\subsubsection{ The kinetic energy terms }
An important property of the nonlinear realization of chiral symmetry
is that all terms of the model-Lagrangian only have to be invariant
under $SU(3)_V$ transformations in order to ensure chiral
symmetry. This vector transformation depends in general on the
pseudoscalar mesons and thus is local. Covariant derivatives have
to be introduced for the kinetic terms in order to preserve chiral
invariance \cite{paper3}.
The covariant derivative used in this case, reads:
$ D_\mu = \pam\, + [\Gamma_\mu,\,\,]$
with $\Gamma_\mu=-\fr{i}{2}[u^\dg\pam u + u\pam u^\dg]$, where
$u=\exp\Bigg[\fr{i}{\sigma_0}\pi^a\lambda^a\gamma_5\Bigg]$ is the unitary
transformation operator. The pseudoscalar mesons are given as parameters
of the symmetry transformation. \\
The kinetic energy terms read
\bea
\label{kinetic}
{\cal L}_{kin} &=& i\Tr \overline{B} \gamma_{\mu} D^{\mu}B
                + \frac{1}{2} \Tr D_{\mu} X D^{\mu} X
+  \Tr (u_{\mu} X u^{\mu}X +X u_{\mu} u^{\mu} X)
                + \frac{1}{2}\Tr D_{\mu} Y D^{\mu} Y \nonumber \\
               &+&\frac {1}{2} D_{\mu} \chi D^{\mu} \chi
                - \frac{ 1 }{ 4 } \Tr
\left(\tilde V_{ \mu \nu } \tilde V^{\mu \nu }  \right)
- \frac{ 1 }{ 4 } \Tr \left(F_{ \mu \nu } F^{\mu \nu }  \right)
- \frac{ 1 }{ 4 } \Tr \left( {\cal A}_{ \mu \nu } {\cal A}^{\mu \nu }
 \right)\, .
\eea
In (\ref{kinetic}) $B$ is the baryon octet, $X$ the scalar meson
multiplet, $Y$ the pseudoscalar chiral singlet, $\tilde{V}^\mu$ (${\cal
A}^\mu$) the renormalised vector (axial vector) meson multiplet with
the field strength tensor
$\tilde{V}_{\mu\nu}=\pa_\mu\tilde{V}_\nu-\pa_\nu\tilde{V}_\mu$ $({\cal
A}_{\mu\nu}= \pa_\mu{\cal A}_\nu-\pa_\nu{\cal A}_\mu $).  $F_{\mu\nu}$
is the field strength tensor of the photon and $\chi$
is the scalar, iso-scalar dilaton (glueball) -field.

\subsubsection{Baryon-meson interaction}
Except for the difference in Lorentz indices, the SU(3) structure of the
baryon -meson interaction terms are the same for all mesons.
This interaction for a general meson field $W$ has the form
\be
{\cal L}_{BW} =
-\sqrt{2}g_8^W \left(\alpha_W[\ovl{B}{\cal O}BW]_F+ (1-\alpha_W)
[\ovl{B} {\cal O}B W]_D \right)
- g_1^W \frac{1}{\sqrt{3}} \Tr(\ovl{B}{\cal O} B)\Tr W  \, ,
\ee
with $[\ovl{B}{\cal O}BW]_F:=\Tr(\ovl{B}{\cal O}WB-\ovl{B}{\cal O}BW)$ and
$[\ovl{B}{\cal O}BW]_D:= \Tr(\ovl{B}{\cal O}WB+\ovl{B}{\cal O}BW)
- \frac{2}{3}\Tr (\ovl{B}{\cal O} B) \Tr W$.
The different terms -- to be considered -- are those for the interaction
of baryons  with
scalar mesons ($W=X, {\cal O}=1$), with
vector mesons  ($W=\tilde V_{\mu}, {\cal O}=\gamma_{\mu}$ for the vector and
$W=\tilde V_{\mu \nu}, {\cal O}=\sigma^{\mu \nu}$ for the tensor
interaction),
with axial vector mesons ($W={\cal A}_\mu, {\cal O}=\gamma_\mu \gamma_5$)
and with
pseudoscalar mesons ($W=u_{\mu},{\cal O}=\gamma_{\mu}\gamma_5$), respectively.
For the current investigation the following interactions are relevant:
Baryon-scalar meson interactions generate the baryon masses through
coupling of the baryons to the non-strange $ \sigma (\sim
\langle\bar{u}u + \bar{d}d\rangle) $ and the strange $
\zeta(\sim\langle\bar{s}s\rangle) $ scalar quark condensate.  After
insertion of the scalar meson matrix $X$, one obtains the baryon masses
as
\bea
m_N &=& m_0 - \fr{1}{3}g^S_8(4\alpha_S-1)(\sqrt{2}\zeta - \sigma)  \nonumber  \\
m_{\Lambda} &=& m_0 - \fr{2}{3}g^S_8(\alpha_S-1)(\sqrt{2}\zeta - \sigma)  \nonumber  \\
m_{\Sigma} &=& m_0 + \fr{2}{3}g^S_8(\alpha_S-1)(\sqrt{2}\zeta - \sigma) \\
m_{\Xi} &=& m_0 + \fr{1}{3}g^S_8(2\alpha_S + 1)(\sqrt{2}\zeta - \sigma)  \nonumber
\eea
with $m_0=g^S_1(\sqrt{2}\sigma + \zeta)/\sqrt{3}$.
The parameters $ g_1^S , g_8^S $ and $ \alpha_S $ can be used
to fix the baryon masses to their experimentally measured
vacuum values. It should be emphasised that the nucleon mass
also depends on the {\em strange condensate} $ \zeta $.
Recently, the vector meson properties were investigated in nuclear matter
for the special case of $\alpha_S=1$ and $g_1^S=\sqrt 6 g_8^S$ \cite{hartree}.
Then the nucleon mass depends only on the non--strange quark condensate.
In the present investigation, the general case will be used
to study hot and strange hadronic matter \cite{kristof1} and
takes into account the baryon couplings to both scalar fields ($\sigma$
and $\zeta$) while summing over the baryonic tadpole diagrams to
investigate the effect from the baryonic Dirac sea in the relativistic
Hartree approximation \cite{kristof1}.

In analogy to the baryon-scalar meson coupling there exist two
independent baryon-vector meson interaction terms corresponding
to the F-type (antisymmetric) and $D$-type (symmetric) couplings.
Here we will use the symmetric coupling because -- from
the universality principle  \cite{saku69} and
the vector meson dominance model -- one can conclude
that the antisymmetric coupling should be small.
We realize it by setting $\alpha_V=1$
for all fits. Additionally we decouple the strange vector field
$ \phi_\mu\sim\bar{s}
\gamma_\mu s $ from the nucleon by setting $ g_1^V=\sqrt{6}g_8^V $.
The remaining baryon-vector meson interaction reads
\be
{\cal L}_{BV}=-\sqrt{2}g_8^V\Big\{[\bar{B}\gamma_\mu BV^\mu]_F+\Tr\big(\bar{B}\gamma_\mu B\big)
\Tr V^\mu\Big\}\, .
\ee

\subsubsection{Meson-meson interactions}
The Lagrangian describing the interaction for the scalar mesons, $X$,
and pseudoscalar singlet, $Y$, is given as \cite{paper3}
\bea
\label{cpot}
{\cal L}_0 &= &  -\frac{ 1 }{ 2 } k_0 \chi^2 I_2
     + k_1 (I_2)^2 + k_2 I_4 +2 k_3 \chi I_3,
\eea
with $I_2= \Tr (X+iY)^2$, $I_3=\det (X+iY)$ and $I_4 = \Tr (X+iY)^4$.
In the above, $\chi$ is the scalar color singlet gluon field. It is
introduced in order to satisfy the QCD trace anomaly, i.e. the nonvanishing
energy-momentum tensor $\Theta_\mu^\mu = (\beta_{QCD}/2g)\langle
G^a_{\mu\nu}G^{a,\mu\nu}\rangle$, where $G^a_{\mu\nu}$ is the gluon
field tensor.  \\
A scale breaking potential
\be
\label{lscale}
  {\cal L}_{\mathrm{scalebreak}}=- \frac{1}{4}\chi^4 \ln
   \frac{ \chi^4 }{ \chi_0^4}
 +\frac{\delta}{3}\chi^4 \ln \frac{I_3}{\det \langle X \rangle_0}
\ee
is introduced additionally and yields
\be
\theta_\mu^\mu=4{\cal L}-\chi\fr{\pa{\cal L}}{\pa\chi}-2\pa_\mu\chi
\fr{\pa{\cal L}}{\pa(\pa_\mu\chi)} = \chi^4 .
\ee
It allows for the identification of the $\chi$ field width the gluon
condensate $\Theta_\mu^\mu=(1-\delta)\theta_\mu^\mu=(1-\delta)\chi^4$.
Finally the term
${\cal L}_{\chi} = - k_4 \chi^4 $
generates a phenomenologically consistent finite vacuum expectation
value. We shall use the frozen glueball approximation i.e. assume
$\chi = \langle 0|\chi|0\rangle\equiv\chi_0$, since the variation of
$\chi$ in the medium is rather small \cite{paper3}.

The Lagrangian for the vector meson interaction is written as
\bea
{\cal L}_{vec} &=&
    \fr{m_V^2}{2}\fr{\chi^2}{\chi_0^2}\Tr\big(\tilde{V}_\mu\tilde{V}^\mu\big)
+   \fr{\mu}{4}\Tr\big(\tilde{V}_{\mu\nu}\tilde{V}^{\mu\nu}X^2\big) 
+ \fr{\lambda_V}{12}\Big(\Tr\big(\tilde{V}^{\mu\nu}\big)\Big)^2 +
    2(\tilde{g}_4)^4\Tr\big(\tilde{V}_\mu\tilde{V}^\mu\big)^2  \, .
\eea
The vector meson fields, $\tilde{V}_\mu$ are related to the
renormalized fields by
$V_\mu = Z_V^{1/2}\tilde{V}_\mu$, with $V = \omega, \rho, \phi \, $.
The masses of $\omega,\rho$ and $\phi$ are fitted from $m_V, \mu$ and
$\lambda_V$. 
%
\subsubsection{Explicit chiral symmetry breaking}
The explicit symmetry breaking term is given as \cite{paper3}
\be
 {\cal L}_{SB}=\Tr A_p\left(u(X+iY)u+u^\dagger(X-iY)u^\dagger\right)
\label{esb-gl}
\ee
with $A_p=1/\sqrt{2}{\mathrm{diag}}(m_{\pi}^2 f_{\pi},m_\pi^2 f_\pi, 2 m_K^2 f_K
-m_{\pi}^2 f_\pi)$ and $m_{\pi}=139$ MeV, $m_K=498$ MeV. This
choice for $A_p$, together with the constraints
$\sigma_0=-f_\pi$, $\zeta_0=-\frac {1}{\sqrt 2} (2 f_K -f_\pi)$
on the VEV on the scalar condensates assure that
the PCAC-relations of the pion and kaon are fulfilled.
With $f_{\pi} = 93.3$~MeV and $f_K = 122$~MeV we obtain $|\sigma_0| =
93.3$~MeV and $|\zeta_0 |= 106.56$~MeV.

\subsection{Mean field approximation}

We next proceed to study the hadronic properties in the chiral SU(3) model.
The Lagrangian density in the mean field approximation is given as
\begin{eqnarray}
{\cal L}_{BX}+{\cal L}_{BV} &=& -\sum_i\overline{\psi_{i}}\, [g_{i
\omega}\gamma_0 \omega + g_{i\phi}\gamma_0 \phi
+m_i^{\ast} ]\,\psi_{i} \\
{\cal L}_{vec} &=& \frac{1}{2}m_{\omega}^{2}\frac{\chi^2}{\chi_0^2}\omega^
2+g_4^4 \omega^4 +
\frac{1}{2}m_{\phi}^{2}\frac{\chi^2}{\chi_0^2}\phi^2+g_4^4
\left(\fr{Z_\phi}{Z_\omega}\right)^2\phi^4\\
{\cal V}_0 &=& \frac{ 1 }{ 2 } k_0 \chi^2 (\sigma^2+\zeta^2)
- k_1 (\sigma^2+\zeta^2)^2
     - k_2 ( \frac{ \sigma^4}{ 2 } + \zeta^4)
     - k_3 \chi \sigma^2 \zeta \nonumber \\
&+& k_4 \chi^4 + \frac{1}{4}\chi^4 \ln \frac{ \chi^4 }{ \chi_0^4}
 -\frac{\delta}{3} \chi^4 \ln \frac{\sigma^2\zeta}{\sigma_0^2 \zeta_0} \\
{\cal V}_{SB} &=& \left(\frac{\chi}{\chi_0}\right)^{2}\left[m_{\pi}^2 f_{\pi}
\sigma
+ (\sqrt{2}m_K^2 f_K - \frac{ 1 }{ \sqrt{2} } m_{\pi}^2 f_{\pi})\zeta
\right],
\end{eqnarray}
where $m_i^* = -g_{\sigma i}{\sigma}-g_{\zeta i}{\zeta} $ is the
effective mass of the baryon of type i ($i = N,\Si ,\La ,\Xi$).
In the above, $g_4=\sqrt {Z_\omega} \tilde g_4$ is the renormalised
coupling for $\omega$-field.
The thermodynamical potential of the grand
canonical ensemble, $\Omega$, per unit volume $V$ at given chemical
potential $\mu$ and temperature $T$ can be written as
\bea
\frac{\Omega}{V} &=& -{\cal L}_{vec} - {\cal L}_0 - {\cal L}_{SB}
- {\cal V}_{vac} + \sum_i\frac{\gamma_i }{(2 \pi)^3}
\int d^3k\,
E^{\ast}_i(k)\Big(f_i(k)+\bar{f}_i(k)
\Big) \\ \nonumber
&&- \sum_i\frac{\gamma_i }{(2 \pi)^3}\,\mu^{\ast}_i
\int d^3k\,\Big(f_i(k)-\bar{f}_i(k)\Big)\, .
\label{OmegaV}
\eea
Here the vacuum energy (the potential at $\rho=0$) has been subtracted
in order to get a vanishing vacuum energy. In (\ref{OmegaV}) $\gamma_i$
are the spin-isospin degeneracy factors.  The $f_i$ and $\bar{f}_i$ are
thermal distribution functions for the baryon of species, $i$ given in
terms of the effective single particle energy, $E^\ast_i$, and chemical
potential, $\mu^\ast_i$, as
\bea
f_i(k) &=& \fr{1}{{\rm e}^{\beta (E^{\ast}_i(k)-\mu^{\ast}_i)}+1}\quad ,\quad
\bar{f}_i(k)=\fr{1}{{\rm e}^{\beta (E^{\ast}_i(k)+\mu^{\ast}_i)}+1}, \no \\
\eea
with $E^{\ast}_i(k) = \sqrt{k_i^2+{m^\ast_i}^2}$ and $ \mu^{\ast}_i
                = \mu_i-g_{i\omega}\omega$.
The mesonic field equations are determined by minimizing the
thermodynamical potential \cite{hartree,kristof1}.
These are expressed in terms of 
the scalar and vector densities
for the baryons at finite temperature
\bea
\rho^s_i = \gamma_i
\int \frac{d^3 k}{(2 \pi)^3} \,\frac{m_i^{\ast}}{E^{\ast}_i}\,
\left(f_i(k) + \bar{f}_i(k)\right) \, ; \;\;
\rho_i = \gamma_i \int \frac{d^3 k}{(2 \pi)^3}\,\left(f_i(k) -
\bar{f}_i(k)\right) \,.
\label{dens}
\eea
The energy density and the pressure are given as,
$\epsilon = \Omega/V+\mu_i\rho_i $+TS and $ p = -\Omega/V $.

\subsection{Relativistic Hartree approximation}
%
%
The relativistic Hartree approximation takes into account the effects from
the Dirac sea by summing over the baryonic tadpole diagrams and
the interacting propagator for a baryon of type $i$ has the form
\cite{vacpol}
\begin{eqnarray}
 && G_i^H(p) = \left(\gamma^\mu\bar{p}_\mu+m_i^\ast\right)
\Bigg[\frac{1}{\bar{p}^2 -{m_i^\ast}^2+i\epsilon}\nonumber\\
&+&\frac{\pi i}{E_i^\ast(p)}\left\{\frac{\delta(\bar{p}^0
-E_i^\ast(p))}{{\rm e}^{\beta(E_i^\ast(p)-\mu_i^\ast)}+1}
+ \frac{\delta(\bar{p}^0+E_i^\ast(p))}{
{\rm e}^{\beta(E_i^\ast(p)+\mu_i^\ast)}+1} \right\}\Bigg]\nonumber \\
&\equiv & G_i^F(p) + G_i^D(p),
\end{eqnarray}
where $E_i^\ast(p)=\sqrt{{\bf p}^2+{m_i^\ast}^2}$, $\bar{p}=p+\Sigma_i^V$
and $m_i^\ast=m_i+\Sigma_i^S$. $\Sigma_i^V$ and $\Sigma_i^S$
are the vector and scalar self energies of baryon, $i$ respectively.
In the present investigation (for the study of hot hyperonic matter)
the baryons couple to both the non-strange ($\sigma$) and strange
($\zeta$) scalar fields, so that we have
\be
\Sigma^S_i=-(g_{\sigma i}\tilde{\sigma}+g_{\zeta
i}\tilde{\zeta})\, ,
\ee
where $\tilde{\sigma}=\sigma-\sigma_0$, $\tilde{\zeta}=\zeta-\zeta_0$.
The scalar self-energy $\Sigma^S_i$ can be written
\bea
\Sigma^S_i= i\left(\fr{g_{\sigma i}^2}{m_\sigma^2}
 + \fr{g_{\zeta i}^2}{m_\zeta^2}\right)\int\fr{\rm{d}^4p}{(2\pi)^4}
 \Tr\big[G_i^F(p)+G_i^D(p)\big]e^{ip^0\eta}
\equiv \big(\Sigma^S_i\big)^F + \big(\Sigma^S_i\big)^D \, .
\eea
$ (\Sigma^S_i)^D $ is the density dependent part and is identical
to the mean field contribution
\be
\big(\Sigma^S_i\big)^D = -\left(\fr{g_{\sigma i}^2}{m_\sigma^2}
+ \fr{g_{\zeta i}^2}{m_\zeta^2}\right)\rho_i^s,
\label{denpart}
\ee
with $\rho_i^s$ as defined in (\ref{dens}). 
The Feynman part $ (\Sigma^S_i)^F $ of the scalar part of the self-energy
is divergent. We carry out a dimensional regularization to extract
the convergent part.
Adding the counter terms \cite{kristof1}
\be
\label{ctc}
\left(\Sigma^S_i\right)_{CTC} = - \left(\fr{g_{\sigma i}^2}{m_\sigma^2}
+\fr{g_{\zeta i}^2}{m_\zeta^2}\right)\sum_{n=0}^3\fr{1}{n!}
(g_{\sigma i}\tilde{\sigma}+g_{\zeta i}\tilde{\zeta})^n\beta_{n+1}^{i}\, ,
\ee
yields the additional contribution from the Dirac sea to 
the baryon self energy \cite{kristof1}.
The field equations for the scalar meson fields are then modified to
\be
\frac{\partial(\Omega/V)}{\partial\Phi}\Bigg|_{RHA} =
\frac{\partial(\Omega/V)}{\partial\Phi}\Bigg|_{MFT}
+\sum_i\frac{\pa m_i^\ast}{\pa\Phi}\Delta \rho^s_i = 0
\quad\mbox{with}\quad \Phi = \sigma, \zeta\, ,
\ee
where the additional contribution to the nucleon scalar density is
given as \cite{kristof1}
\be
\Delta\rho^s_i = -\fr{\gamma_i}{4\pi^2} \left[ {m_i^\ast}^3\ln\left(
                  \fr{m_i^\ast}{m_i}\right)
+ m_i^2(m_i-m_i^\ast) - \fr{5}{2}m_i(m_i-m_i^\ast)^2 + \fr{11}{6}
  (m_i-m_i^\ast)^3\right].
\ee

\section{$D$-meson mass modification in the medium}
\label{dmeson}

We now examine the medium modification for the $D$-meson mass
in the hot and dense hadronic matter. In the last section,
the SU(3) chiral model was used to study the hadronic properties
in the medium within the relativistic Hartree approximation.
We assume that the additional effect of charmed particles in the medium
leads to only marginal modifications \cite{roeder} of these hadronic
properties and do not need to be taken into account here.
However, to investigate the medium modification of the $D$-meson mass,
we need to know the interactions of the $D$-mesons with the light
hadron sector.

The light quark condensate has been shown to
play an important role for the shift in the $D$-meson mass in
the QCD sum rule calculations \cite{arata}. In the present chiral model,
the interactions to the scalar fields (nonstrange, $\sigma$ and strange,
$\zeta$) as well as a vectorial interaction and a $\omega$- exchange
term modify the masses for D$^\pm$ mesons in the medium.
These interactions were considered within the SU(3) chiral model
to investigate the modifications of K-mesons in thermal
medium \cite{kmeson}. The scalar meson exchange gives an attractive
interaction leading to a drop of the D (K) -meson masses similar
to a scalar sigma term in the chiral perturbation theory \cite{kaplan}.
In fact, the sigma term corresponding to the scalar kaon-nucleon
\cite{kmeson} (as well as $\pi N$ sigma term) and 
$D$-nucleon attractive interaction
are predicted in our approach automatically by using SU(3) and SU(4)
symmetry, respectively. The pion-nucleon and kaon-nucleon sigma terms
as calculated  from the scalar meson exchange interaction of
our Lagrangian are around 28 MeV and 463 MeV respectively. The value
for KN sigma term calculated in our model is close
to the value of $\Sigma_{KN}$=450 MeV found by lattice gauge calculations 
of \cite{ksgn}. The value for the DN sigma term
within our chiral model turns out to be $\Sigma_{DN}=7366$~MeV 
ignoring the contribution of the charm condensate to  
the nuclear scalar density. In the present investigation
we also consider the effect of repulsive scalar contributions
($\sim (\partial _\mu D^+)(\partial ^\mu D^-)$)  which contribute
in the same order as the attractive sigma term in chiral perturbation
theory.

To consider the medium effect on the $D$-meson masses we generalize the
chiral $SU(3)$-flavor model to include the charmed mesons. The scalar
meson multiplet has now the expectation value
\begin{equation}
\langle X \rangle 
= \left(
\begin{array}{cccc} \sigma/\sqrt 2 & 0 & 0 & 0\\
 0 & \sigma/\sqrt 2 & 0 & 0 \\
 0 &  0 & \zeta & 0 \\
 0 &  0 & 0 & \zeta_c \\
\end{array}\right),
\end{equation}
with $\zeta_c$ corresponding to the $\bar c c$ condensate.
The pseudoscalar meson field P can be written, including the charmed
mesons, as
\begin{equation}
P = \left(
\begin{array}{cccc} \pi^0/\sqrt 2 & \pi^+ & \frac{2 K^+}{1+w} & 0 \\
\pi^- & -\pi^0/\sqrt 2 & 0 & \frac {2 D^-}{1+w_c} \\
\frac {2 K^-}{1+w} & 0 & 0 & 0
\\ 0 & \frac {2 D^+}{1+w_c} & 0 & 0 \end{array}\right),
\end{equation}
where $w=\sqrt 2 \zeta/\sigma$ and  $w_c=\sqrt 2 \zeta_c/\sigma$.
From PCAC, one gets the decay constants for the pseudoscalar mesons
as $f_\pi=-\sigma$, $f_K=-(\sigma +\sqrt 2 \zeta )/2$ and
$f_D=-(\sigma +\sqrt 2 \zeta_c )/2$. In the present calculations,
the value for the D-decay constant will be taken to be 135 MeV
\cite{weise}. We note that for the decay constant of $D_s^+$,
the Particle Data Group \cite{pdg} quotes a value of 
$f_{D_s^+} \simeq 200 MeV$. Taking a similar value also for
$f_D$ would not affect our results qualitatively, however
(see also \cite{lat03}).
The vector meson interaction with
the pseudoscalar mesons, which modifies the masses of the K(D) mesons,
is given as \cite {kmeson}

\begin{equation}
{\cal L} _ {VP}= -\frac{m_V^2}{2g_V} {\rm {Tr}} (\Gamma_\mu V^\mu) + h.c.
\label{lvp}
\end{equation}
The vector meson multiplet is given as 
$V = {\rm  diag}\big ((\omega +\rho_0)/\sqrt 2,\;
(\omega -\rho_0)/\sqrt 2,\; \phi, \; J/\Psi\big )$. The non-diagonal
components in the multiplet have not been written down 
as they are not relevant for the present investigation.
With the interaction (\ref{lvp}),
the coupling of the $D$-meson to the $\omega$-meson is related to  the
pion-rho coupling as
$g_{\omega D}/g_{\rho \pi \pi}=f_\pi^2 /(2f_D^2)$.

The scalar meson exchange interaction term, which is attractive
for the $D$-mesons, is given from the explicit symmetry breaking term
by equation (\ref {esb-gl}), where $A_p =1/\sqrt 2$
diag ($m_\pi^2 f_\pi$, $m_\pi^2 f_\pi$, 2 $m_K^2 f_K -m_\pi^2 f_\pi$,
 2 $m_D^2 f_D -m_\pi^2 f_\pi$).

The interaction Lagrangian modifying the $D$-meson mass can be written
as \cite{kmeson}
\begin{eqnarray}
\cal L _{D} & = & -\frac {3i}{8 f_D^2} \bar N \gamma^\mu N
( D^+ \partial_\mu D ^- - \partial_\mu D ^+ D^-)\nonumber \\
 &+ & \frac{m_D^2}{2f_D} (\sigma +\sqrt 2 \zeta_c) D^+ D^-
-i g_{\omega D} ( D^+ \partial_\mu D ^- - \partial_\mu D ^+ D^-)
\omega ^ \mu \nonumber \\
& - & \frac {1}{f_D} (\sigma +\sqrt 2 \zeta_c)
(\partial _\mu D^+)(\partial ^\mu D^-)
+\frac {d_1}{2 f_D^2}(\bar N N)
(\partial _\mu D^+)(\partial ^\mu D^-).
\label{lagd}
\end{eqnarray}
In (\ref{lagd}) the first term is the vectorial interaction term
obtained from the first term in (\ref{kinetic}). The second term, which gives
an attractive interaction for the $D$-mesons, is obtained from the
explicit symmetry breaking term (\ref{esb-gl}). The third term,
referring to the interaction in terms of $\omega$-meson exchange,
is attractive for the $D^+$ and repulsive for $D^-$.
The fourth term arises within the present chiral model from
the kinetic term of the pseudoscalar mesons given by the third term
in equation (\ref{kinetic}), when the scalar fields in one of the meson
multiplets, $X$ are replaced by their vacuum expectation values.
The fifth term in (\ref{lagd}) has been written down 
for the $DN$ interactions, analogous to a term of the type
\begin{equation}
{\cal L }_{\tilde D }^{BM} =d_1 Tr (u_\mu u ^\mu \bar B B),
\label{dtld}
\end{equation}
in the SU(3) chiral model. The last two terms in (\ref{lagd})
represent the range term in the chiral model. 

\subsection{Fitting to KN scattering data}

The term (\ref{dtld}) reduces for the $KN$ interactions to
\begin{equation}
{\cal L }_{\tilde D }^{KN} =d_1 \frac{1}{2 f_K^2}
(\bar N N) (\partial _\mu K^+) (\partial ^\mu K^-).
\end{equation}
The coefficient, $d_1$ in the above shall be determined 
as consistent with the $KN$ scattering data \cite{thorsson,barnes,juergen}. 
The isospin averaged $KN$ scattering length 
\be
\bar a _{KN}= \frac {1}{4} (3 a_{KN}^{I=1}+ a_{KN}^{I=0})
\ee
can be calculated to be
\bea
\bar a _{KN} &=&\frac {m_K}{4\pi (1+m_K/m_N)} \Big [ 
-\big(\frac {m_K}{2 f_K} \big )\cdot \frac {g_{\sigma N}}{m_\sigma^2}
-\big(\frac {\sqrt 2 m_K}{2 f_K} \big )\cdot \frac {g_{\zeta N}}{m_\zeta^2}
\nonumber \\
 & - & \frac {2 g_{\omega K} g_{\omega N}}{m_\omega ^2}
-\frac {3}{4 f_K^2} + \frac {d_1 m_K}{2 f_K^2} \Big ].
\eea
Choosing the empirical value of the isospin averaged scattering length
\cite{thorsson,juergen},
\be
\bar a _{KN} \approx -0.255 ~ \rm {fm}
\label{aknemp}
\ee
determines the value for the coefficient $d_1$. 
The present calculations corresponds to the values
$g_{\sigma N}=10.618,\;\; {\rm and}\;\; g_{\zeta N}=-0.4836$ as
consistent with the vacuum baryon masses,
and the other parameters as fitted from the nuclear matter
saturation properties as listed in Ref. \cite{kristof1}.
We consider the case when a quartic vector interaction is present.
The coefficient $d_1$ is evaluated in the mean field and RHA cases 
as ${5.63}/{m_K}$ and ${4.33}/{m_K}$ respectively. 
The contribution from this term is thus seen to be attractive,
contrary to the other term proportional to 
$(\partial_\mu D^+)(\partial ^\mu D^-)$ in (\ref{lagd}) 
which is repulsive. 

\subsection{Chiral perturbation theory}

The effective Lagrangian obtained from chiral perturbation theory
\cite{kaplan} has been used extensively in the literature for the study
of kaons in dense matter. This has a vector interaction
(called the Tomozawa-Weinberg term) as the leading term.
At sub-leading order there are the attractive scalar nucleon
interaction term (the sigma term) \cite{kaplan} as well as
the repulsive scalar contribution (proportional to the kinetic
term of the pesudoscalar meson).
We generalize such an interaction to SU(4) to write down the interaction
of a $D$-meson with a nucleon as
\begin {eqnarray}
{\cal L}_ {DN} & = &
-\frac {3i}{8 f_D^2} \bar N \gamma^\mu N
( D^+ \partial_\mu D ^- - \partial_\mu D ^+ D^-)
+ \frac{\Sigma_{DN}}{f_D^2} (\bar N N ) D^+ D^-\nonumber \\
&+ & \frac {\tilde D}{f_D^2}
(\bar N N) (\partial _\mu D^+) (\partial ^\mu D^-).
\label{ldcpt}
\end{eqnarray}
where $\Sigma _{DN}=\frac {m_d+m_c}{2} \langle N | (\bar d d + \bar c c) |
N \rangle$ in analogy to the definition of $\Sigma_{KN}
=\frac{\bar m +m_s}{2} \langle N | (\bar u u + \bar s s) |
N \rangle$ \cite{ksgn}.
Neglecting the charm
condensate inside the nucleon, this is directly related to the
pion-nucleon sigma term given as $\Sigma _{\pi N}=\bar m
\langle N | (\bar u u + \bar d d)|N\rangle$, with
$\langle N | \bar u u | N \rangle =\langle N | \bar d d | N \rangle$.
In our calculations, we take $m_c=1.3~GeV$, $\bar m=(m_u+m_d)/2=7 MeV$,
and $\Sigma_{\pi N}=45 MeV$,
which gives the value for $\Sigma_{DN}= 2089 ~MeV$.

The last term of the Lagrangian (\ref{ldcpt}) is repulsive and is
of the same order as the attractive sigma term. This, to a large
extent, compensates the scalar attraction due to the scalar
$\Sigma$- term. We fix the coefficient $\tilde D$ from
the $KN$ scattering data \cite{thorsson}.
This involves choosing a value for $\Sigma_{KN}$,
which depends on the strange condensate content of the nucleon.
Its value has, however, a large uncertainty.
We consider the two extreme choices:
$\Sigma _{KN}= 2 m_\pi$ and $\Sigma_{K N}= 450$~MeV.
The coefficient, $\tilde {D}$ as fitted to the empirical
value of the KN scattering length (\ref{aknemp}) is:
\begin{equation}
\tilde D \approx 0.33/m_K - \Sigma_{KN} /m_K^2.
\end{equation}

In the next section, we shall discuss the results for 
the $D$-meson mass modification obtained in the present effective 
chiral model as compared to that using the interaction Lagrangian of 
chiral perturbation theory as well as from other approaches.

\section{Results and Discussions}
\label{results}
To study the $D$-meson masses in  hot and dense hadronic medium
due to its interactions with the light hadrons,
we have generalized the chiral SU(3) model
used for the study of the hot and strange hadronic matter to SU(4)
for the meson sector.
The contributions from the various terms of the interaction Lagrangian
(\ref{lagd}) are shown in Fig. 1 in mean field approximation.  The
vector interaction (A) as well as the $\omega$ exchange (C) terms
(given by the first and the third terms of equation (\ref{lagd}),
respectively) lead to a drop for the $D^+$ mass, whereas they are
repulsive for the $D^-$.  The scalar meson exchange term (B) is
attractive for both $D^+$ and $D^-$. The first term of the range term
(referred to as (D)) of eq. (\ref{lagd}) is repulsive whereas the
second term has an attractive contribution.
This results in a turn over of the $D$-mass at around 0.4 $\rho_0$
above which the last term in (\ref{lagd}) (attractive) dominates. 
The dominant contributions arise from the scalar exchange (B) and the
term (D) (dominated by $d_1$ term at higher densities), 
which lead to a substantial
drop of D meson mass in the medium. The vector terms (A) and (C)
lead to a further drop of $D^+$ mass, whereas for $D^-$ they compete 
with the contributions from the other two contributions.
The effect from the nucleon Dirac sea on the mass modification of the
$D$-mesons is shown in Fig. 2.  This gives rise to smaller
modifications as compared to the mean field calculations though 
qualitative features remain the same.

\begin{figure}
\begin{center}
\includegraphics[width=16cm,height=10cm]{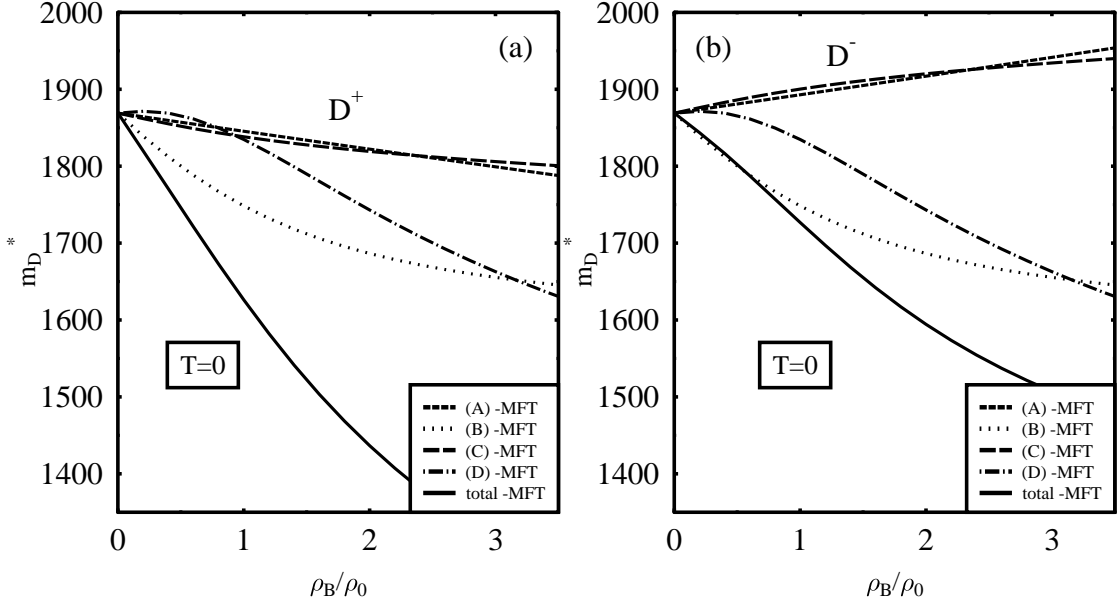}  
\caption{
\label{mdt0imft}
Contributions to the masses of D$^\pm$ mesons due to the various
interactions in the present chiral model in the mean field
approximation. The curves refer to individual contributions from
(A) the vectorial interaction, (B) scalar exchange, (C) $\omega$ exchange,
(D) the last two terms of the equation (\ref{lagd}).
The solid line refers to the total contribution.
}
\end{center}
\end{figure}

\begin{figure}
\begin{center}
\includegraphics[width=16cm,height=10cm]{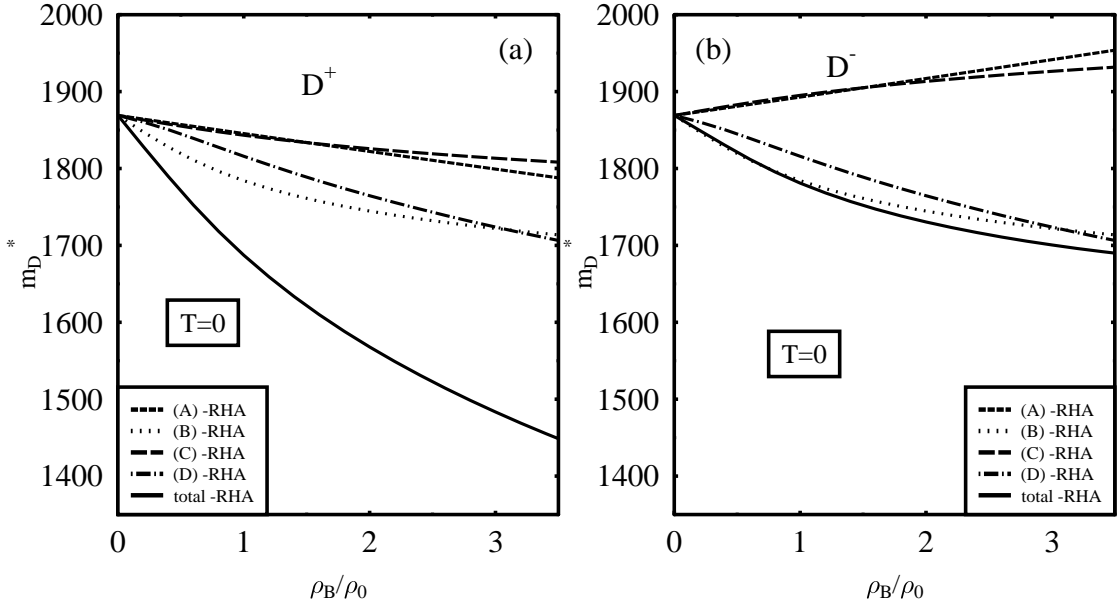}  
\caption{
\label{mdt0}
Same as in figure (\protect\ref{mdt0imft}), but in the relativistic
Hartree approximation. The contributions to the masses of $D^\pm$ mesons
due to the various interactions are seen to be smaller when the
Dirac sea effects are taken into account.}
\end{center}
\end{figure}

\begin{figure}
\begin{center}
\includegraphics[width=16cm,height=10cm]{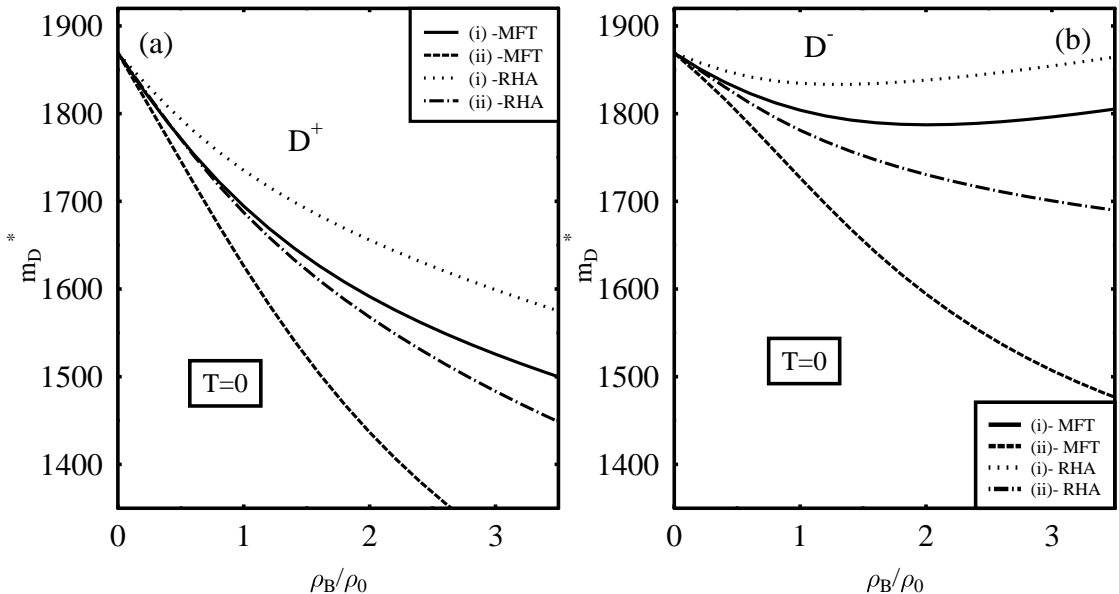}
\caption{
\label{md0}
Masses of $D^\pm$ mesons due to the interactions in the 
chiral effective model at $T=0$.
}
\end{center}
\end{figure}

In Fig. 3 the masses of the $D$-mesons are plotted for $T=0$
in the present chiral model. We first consider the situation 
(case (i)) when the Weinberg-Tomozawa term is supplemented by the
scalar and vector meson exchange interactions \cite{kmeson}.
The other case (ii) corresponds to the inclusion
of the last two terms in (\ref{lagd}) with the parameter
$d_1$ determined from the empirical value
of the scattering length (\ref{aknemp}). 
There is seen to be a substantial drop of D-meson masses due to
the inclusion of the range term.

\begin{figure}
\begin{center}
\includegraphics[width=16cm,height=10cm]{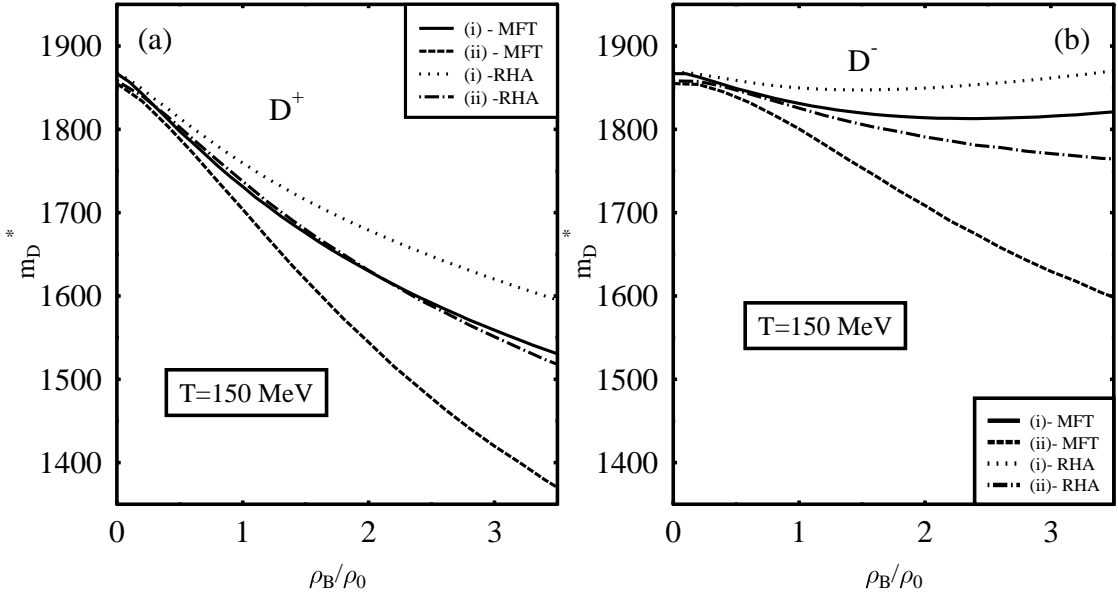}
\caption{
\label{md150}
Masses of $D^\pm$ mesons in the chiral effective model for $T=150$~MeV.
}
\end{center}
\end{figure}

In Fig. \ref{md150}, the masses are plotted for $T=150$~MeV.  One might
note here, that the drop is smaller as compared to at $T=0$ at finite
densities. This is due to the fact that the nucleon mass increases with
temperature at finite densities \cite{liko,kristof1}. Such a behaviour
of the nucleon mass with temperature was also observed earlier within
the Walecka model by Ko and Li \cite {liko} in a mean field
calculation. This subtle behaviour of the baryon self energy, given by
(\ref{denpart}) in the mean field approximation, can be understood in
the following manner:  The scalar self energy (\ref{denpart})
increases due to the thermal distribution functions at finite 
temperatures, whereas at higher temperatures there are also
contributions from higher momenta which lead to lower values of 
self energy. These competing effects give rise to the observed
behaviour of the effective baryon masses with temperature at finite
densities. This increase in the nucleon mass with temperature is also
reflected in the vector meson ($\omega$, $\rho$ and $\phi$) masses
in the medium \cite{kristof1}. However at zero density,
due to effects arising only from the thermal distribution functions, 
the masses are seen to drop with temperature.

\begin{figure}
\begin{center}
\includegraphics[width=16cm,height=10cm]{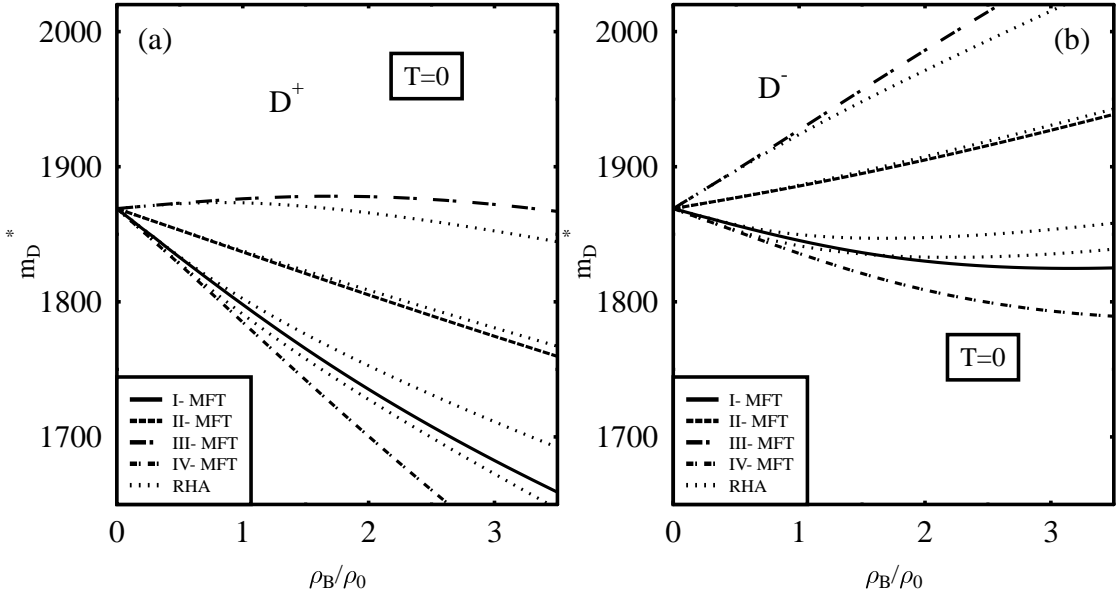}
\caption{
\label{mdt0cpt}
Masses of $D^\pm$ mesons at $T=0$ due to the interactions of
chiral perturbation theory (see text for details).
}
\end{center}
\end{figure}

\begin{figure}
\begin{center}
\includegraphics[width=16cm,height=10cm]{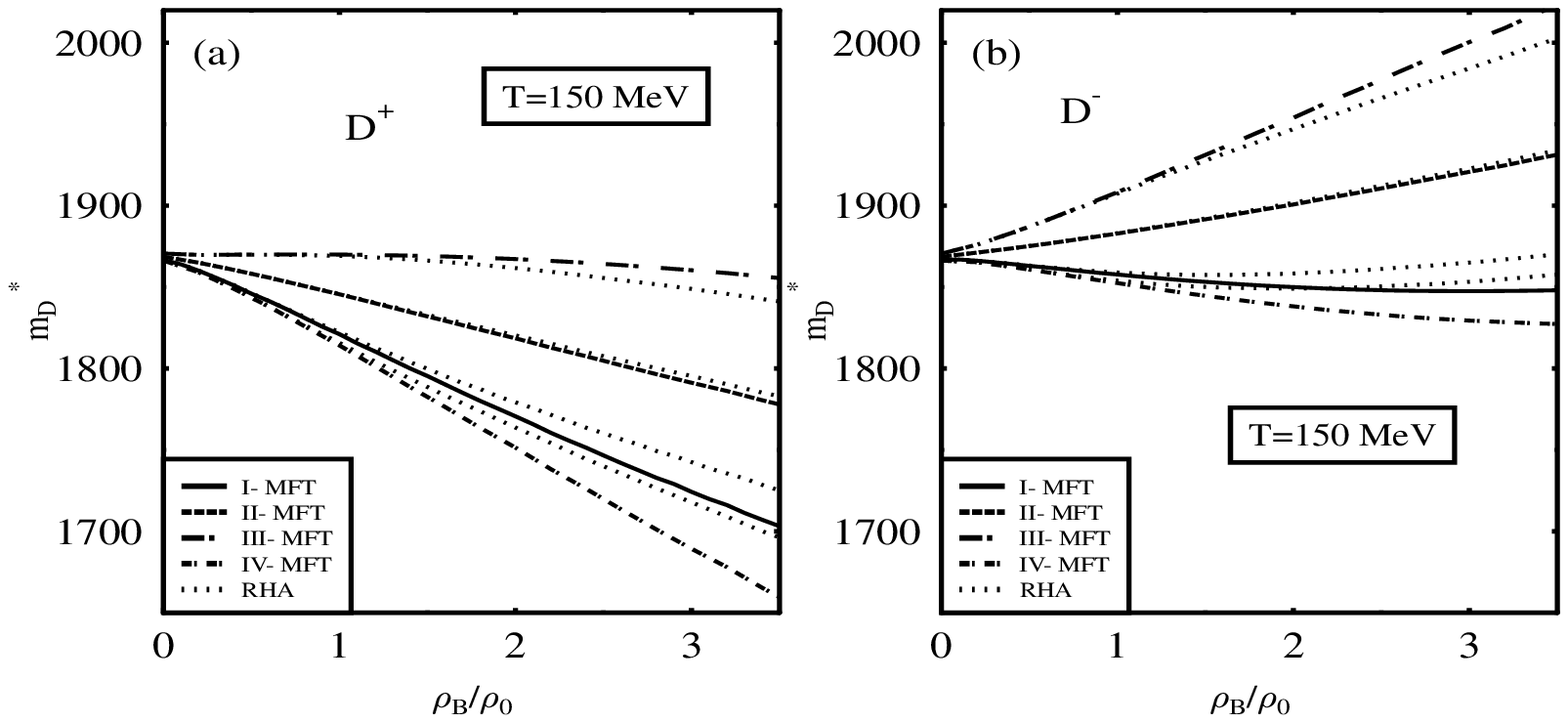}
\caption{
\label{md150cpt}
Masses of D$^\pm$ mesons due to the interactions of
chiral perturbation theory for $T=150$ MeV (see text for details).
}
\end{center}
\end{figure}

We compare the results obtained in the present model to
those using the interaction Lagrangian of
chiral perturbation theory from Ref. \cite{kaplan}. 
The masses obtained from chiral perturbation theory are plotted
in Figs. \ref{mdt0cpt} and \ref{md150cpt} for $T=0$ and 150 MeV.
The $D^\pm$ masses are plotted for the cases: (I) in the
absence of the $(\partial _\mu D^+) (\partial ^\mu D^-)$,
(II) $\tilde D$ corresponding to $\Sigma_{KN}= 2m_\pi$,
that is when the strangeness content of the nucleon is zero,
(III) $\tilde D$ corresponding to $\Sigma_{KN}= 450$ MeV,
(IV) $\tilde D$ corresponding to $\Sigma_{KN}= 450$ MeV
and $\Sigma_{DN}$ =7366 MeV, as calculated from the 
effective chiral model (\ref{lagd}). The case (I) 
shows stronger drop of the $D^+$ mass in the medium 
as compared to (II) and (III) due to the exclusion of the
scalar repulsion term. For $D^-$ however there are 
cancelling  effects from the sigma term and the Weinberg-Tomozawa 
interactions leading to only moderate mass modification. 
The repulsive term  for (III) corresponding to the larger value
of $\tilde D$  has a higher contribution as compared to (II) 
as expected. However, using the value for $\Sigma_{DN}$
as predicted form the chiral effective model, leads to
a larger drop for the D-meson masses from the sigma term
as shown in curve (IV). The relativistic Hartree approximation
gives rise to smaller mass modification as compared to the
mean field case. In the chiral effective model ({\ref{lagd})
however the D$^+$-meson mass has a stronger drop as compared to
the chiral perturbation theory. The value of $d_1$ in (\ref{lagd}) 
as fitted from low energy KN scattering data has a higher value 
to overcome the repulsive $\omega$- exchange term, 
which is absent in chiral perturbation theory. 
This  leads to the range term 
in the present chiral model to be attractive 
of the same in chiral perturbation theory. As a result, the
mass of the $D^+ D^-$ pair experiences a larger drop in the medium
which as we shall see leads to the decay of charmonium states
to such a pair.

When the vectorial interaction -- supplemented by the scalar sigma
term -- is considered (case I), i.e., ignoring the repulsive
terms proportional to $(\partial_\mu D) (\partial ^\mu D)$,
one obtains mass drops for $D^\pm$
to be around 67 MeV and 19 MeV at nuclear matter saturation density.
These values are similar to those obtained in the QCD sum rule
calculations of Ref. \cite{weise}. A similar drop of the D-meson
mass is also predicted by the QMC model \cite{qmc}.
We stress that the present model gives stronger modifications
for the $D$-meson masses than chiral perturbation theory.

It is interesting to compare the behaviour of D and K meson masses 
in a medium. The masses of $K^-$ as well as $D^+$ drop 
in the medium. For the kaons, the vector interaction 
in the chiral perturbation theory 
is the leading contribution giving rise 
to a drop (increase) of the mass of $K^-$ ($K^+$). 
The subleading contributions arise from the sigma and range terms 
with their coefficients as fitted to the KN scattering data 
\cite{barnes}. 
Ignoring the charm condensate contribution 
in the nucleon in the chiral perturbation theory, 
the  $D^-$ mass also increases in the medium 
as in the cases II and III (see figures 5 and 6) similar to the 
behaviour of $K^+$. However, choosing the value for $\Sigma_{DN}$ 
as calculated in the present chiral effective model, 
the mass of $D^-$ drops in the medium (case IV). 
In the chiral effective 
model, the scalar exchange term as well as the range term
(which turns attractive for densities above $0.4 \rho_0$)
lead to the drop of both $D^+$ and $D^-$ masses in the medium.
We note here that a similar behaviour is also obtained for the $K^+$.
Firstly, it increases up to around a density of $0.8 \rho_0$
and then drops due to the range term becoming attractive
at higher densitites. However, though the qualitative features
remain the same, the medium modification for $K^+$ is
much less pronounced as compared to that of the $D^+$ in the medium.
The medium modifications for the kaons and D-mesons
in either model are obtained as consistent with the
low energy KN scattering data. 

\begin{figure}
\begin{center}
\includegraphics[width=16cm,height=10cm]{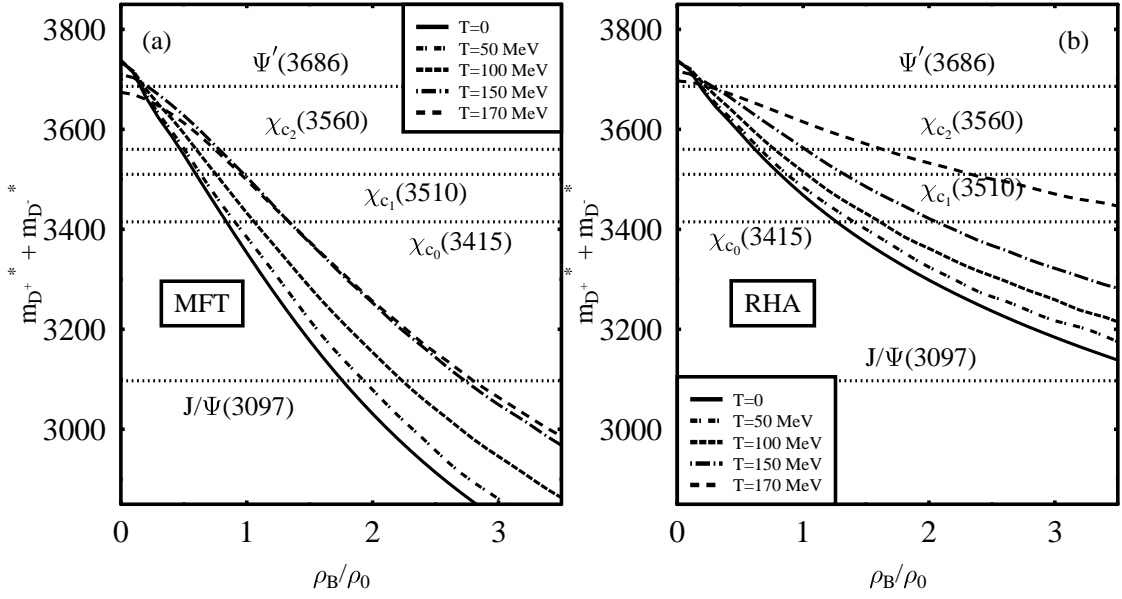}
\caption{
\label{crmb}
The sum of the masses for $D^\pm$-mesons plotted versus density
for different temperatures. The masses for the $\Psi ^\prime$, 
$\chi_{c}$ and $J/\Psi$ are also shown to indicate 
the threshold conditions for the decay of these quarkonium
states into $D \bar D$ pairs.
The relativistic Hartree approximation for a given temperature
is seen to give rise to a higher threshold for the density
as compared to the mean field calculations.
}
\end{center}
\end{figure}

The decay widths of the charmonium states can be modified by the level
crossings between the excited states of J/$\Psi$ (i.e.,
$\Psi^\prime$,$\chi_c$) and the threshold for $D\bar D$ creation due to
the medium modifications of the $D$-meson masses \cite{friman}. In the
vacuum, the resonances above the $D\bar D$ threshold, for example the
$\Psi ''$ state, has a width of 25 MeV due to  the strong open charm
channel. On the other hand, the resonances below the threshold  have a
narrow width of a few hundreds of KeV, only. With the medium
modification of the $D^\pm$-meson masses, the channels for the excited
states of $J/\Psi$, like $\chi_c$, $\Psi^\prime$ decaying to $D^+D^-$
pairs can open up at finite temperatures and densities. 
We show the temperature and density dependence of the mass
of a $D^+ D^-$ pair in Fig. (\ref{crmb}) for the chiral effective Lagrangian. 
The masses of the $J/\Psi$ as well as its excited
states are also shown to indicate the threshold values of temperature
and density when their decay to $D^+ D^-$ pairs becomes accessible. 
Indeed we find that due to the substantial medium modification
of the D-meson (especially of $D^+$ meson) that charmonium states
will dissociate already at quite moderate densities.
There is a strong drop of the mass of $D^+ D^-$ pair
in the mean field calculations, which can lead
to $J/\Psi$ decaying to a pair of D-mesons at around 2-3 times 
the nuclear matter density. The relativistic Hartree approximation 
is seen to give rise to higher threshold values in density. 
At zero density, one sees that the $\Psi^\prime$ decaying 
to $D\bar D$ becomes accessible at a temperature of around 160 MeV
which is around the chiral phase transition.
Studies of charmonium dissociation using heavy quark potential inferred
from lattice data, however, predict smaller values for the
dissociation of $\Psi^\prime$ of around 0.1-0.2 $T_c$ in \cite{digal}
and 0.5 $T_c$ in \cite{wong}. 
In our present model, $\chi_c$ remains stable at zero density,
even up to a temperature of 170 MeV as compared to dissociation
temperatures of $\chi_c$ of around 0.74 $T_c$ in ref. \cite {digal}
and around 0.9 $T_c$ in \cite{wong}. 
The density modifications of the D-meson masses are seen to be
large whereas the mass modifications are seen to be rather 
insensitive to temperature. 

After the level crossings one would
naively expect that the decay widths of $\Psi '$ and $\chi_c$ states
will increase drastically with density. The decay of charmonium states
to $D\bar D$ has been studied in Ref. \cite{brat6,friman}.  It is seen
to depend sensitively on the relative momentum in the final state.
These excited states might  become narrow \cite{friman} though the
$D$-meson mass is decreased appreciably at high temperatures and
densities. It may even vanish at certain momenta corresponding to
nodes in the wavefunction \cite{friman}. Though the decay widths for
these excited states can be modified by their wave functions, the
partial decay width of $\chi_{c2}$, due to absence of any nodes, can
increase monotonically with the drop of the $D^+ D^-$ pair mass in the
medium \cite{friman}. This can give rise to depletion in the $J/\Psi$
yield in heavy ion collisions. The dissociation of the quarkonium
states ($\Psi^\prime$, $\chi_c$, $J/\Psi$) into $D\bar D$ pairs have 
also been studied \cite{digal,wong} by comparing their binding energies 
with lattice results on temperature dependence of heavy quark effective
potential \cite{lattice}. The dissociation occurs since the open charm
mass drops faster with temperature than the mass of the excited 
charmonium \cite{digal}. The medium effects on the charmonium masses
have recently been studied in a perturbative QCD approach \cite{leeko}
which shows an appreciable drop of $\Psi^\prime$ in nuclear matter.
Accounting for the mass modifications of the charmonia will change
the threshold conditions for the decay of these states to 
$D \bar D$ pairs, and, in turn, modify the $J/\Psi$ yield
in the heavy ion collision experiments. It can also have observable 
consequences in the dilepton spectra in the $\bar p A$ annihilation 
experiments \cite{leeko} in the future GSI facility \cite{gsi}.

\section{summary}
To summarize we have investigated in a chiral model the temperature and
density dependence of the $D, \bar D$-meson masses arising from the
interactions with the nucleons and scalar and vector mesons. The
properties of the light hadrons -- as studied in a $SU(3)$ chiral model
-- modify the $D$-meson properties in the hot and dense hadronic
medium. The SU(3) model with paramaters fixed from the properties of
hadron masses, nuclei and hypernuclei and KN scattering data,
is extended in a controlled fashion to SU(4) taking into account
all terms up to the next to leading order arising in chiral perturbative
expansion to derive the interactions of $D$-mesons 
with the light hadron sector. The important advantage of the
present approach is that the DN, KN as well as $\pi N$ sigma terms 
are calculated within the model. The model predictions for
the $\pi N$ and KN sigma terms are reasonable, the value 
for $\Sigma_{KN}$ from the model being in agreement with 
lattice gauge calculations. Using the
Lagrangian from chiral perturbation theory with a vectorial
Tomozawa-Weinberg interaction, supplemented by an attractive scalar
interaction (the sigma term) for the $DN$ interactions, the results
obtained are seen to be similar to earlier finite density calculations
of QCD sum rules \cite{weise} as well as to the quark meson coupling
(QMC) model \cite{qmc}. However, the presence of the repulsive
range term, given by the last term in (\ref{ldcpt}) reduces
the drop in mass. The chiral effective model, which is adjusted to 
describe nuclear properties, dominantly due to the
scalar exchange and the attractive range term 
gives a larger drop of the $D$-meson masses at finite density 
as compared to chiral perturbation theory, 
The effect of the baryon Dirac sea for the hot hyperonic matter 
using the chiral model gives a higher value for the
$D$-meson masses as compared to the mean field calculations. 
The medium modification of the $D$-masses can lead to a suppression in the
$J/\Psi$- yield in heavy-ion collisions. 
In MFT, we find that $J/\Psi$ dissociates into $D\bar D$ pairs
already at 2-3 $\rho_0$. 
The relativistic Hartree approximation gives rise to somewhat 
higher values for the threshold densities for the quarkonium 
decaying to $D^+D^-$ pairs.
The density dependence of D-mass is seen to be the dominant medium 
effect as compared to the temperature dependence. 
The strong density dependence of the D-meson optical potential
can be tested with the existing data on $J/\Psi$ 
production from SPS experiments as well as by the future
GSI facility \cite{gsi} which is currently under investigation. 

\begin{acknowledgements}
We thank J. Reinhardt, I. Shovkovy and A. P. Kostyuk for fruitful
discussions.  One of the authors (AM) is grateful to the Institut f\"ur
Theoretische Physik for warm hospitality. AM acknowledges financial
support from Bundesministerium f\"ur Bildung und Forschung (BMBF) and
ELB to Deutsche Forschungsgemeinschaft (DFG). The support from the 
Frankfurt Center for Scientific Computing (CSC) is gratefully 
acknowledged.

\end{acknowledgements}


\end{document}